# iRECLASS – AN AUTOMATIC SYSTEM FOR RECORDING CLASSES

Edson Lopes
edsolp_at_gmail

José Caetano
caetano.jose_at_gmail

António Abreu
antonio.abreu_at_
estsetubal.ips.pt

Frederico Grilo
frederico.grilo_at
estsetubal.ips.pt

Polytechnic Institute of Setúbal, Setúbal, Portugal

## ABSTRACT

*This paper presents the details of a system capable of recording on video a traditional class. By traditional class it is meant a teacher, a blackboard and a white canvas where course notes are projected. The system is able to track the movements of the lecturer, while recording it on video at the required frame rate (e.g., 25 fps). The system is also capable of understanding five arm gestures made by the lecturer with the intent of controlling which scenario is recorded: himself, the blackboard or the white canvas. The remaining two gestures are for start/stop the recorder. The system is composed by a Kinect sensor, a video camera, a microphone, one pan-tilt system and one pan system, using a total of three step motors.*

## 1. INTRODUCTION

According to (1), lecture caption began circa the year 2000. But only in the past few years we have witnessed a major offer of classes in the internet made available from top player universities, which are now know as Massive Open Online Courses (MOOCs) ((2) and references therein; see (3) for interesting statistics from a MOOC provider). Such classes are offered in two typical forms: a video of a traditional class (i.e., a teacher lecturing before students in a classroom), and a video of lecture notes, where an audio channel of the voice of the lecturer is superimposed.

The latter is economical in what regards technology – it requires an ordinary personal computer and a tablet with a pen; the brilliance or the eloquence of the speaker makes the rest. However, proper video editing is necessary to remove the less good parts. Such editing is typically made by the lecturer. This system is not applicable to any subject/topic. Specifically it is not applicable to situations where learning is facilitated by watching doing.

The former shows more potential with respect to the availability of classes to the general public, because lecturing in a classroom is presently the typical system. All there is to do is to install the recording equipment in the classroom and have a camera operator following the lecturer with a video camera. Also, some video editing is required, which sometimes is made by the camera operator. It is the most expensive system, but as mentioned, the one potentially most able to offer the majority of video classes.

Although such revolution seems to be new, it started while back with the advent of Youtube. As everyone knows, Youtube is for everybody, and the ones gifted to share their knowledge with others, whatever knowledge they have, are better prepared and willing to record their explanations on video.

With the putative massive video production in order, not many will have the means to have at their disposal the good services of a camera operator to record the desired classes/explanations. The possibility of circumventing the necessity of a camera operator is an interesting hypothesis, as it opens the opportunity to a new market of automatic recording systems.

However, circumventing the human factor (in this case the camera operator) is not easy, as he makes important decisions about what scene to record, thus improving the value of the recorded material. The possibilities are: record the lecturer, because we are used to focus attention on the speaker; the white board, where slides are projected; and the blackboard, where the lecturer writes important material. The camera operator can make decisions about what scene to record based on the kind of explanations the lecturer is given at any moment. For instance, the lecturer may be explaining some complex graphic or table, requiring the attention of the students to the slide and not to him. It is not difficult for the camera operator to understand this situation, and thus he decides to aim the camera to the slide and stay there as long as the explanation lasts. However, the most apt person to decide what scene shall be recorded at any given moment is the lecturer itself, thus requiring some ad-hoc form of communication with the camera operator. Natural options are voice and gestures.

The required technology to make video records of a class is mature and not expensive. The ordinary personal computer (e.g., around USD500) is up to the task. The now ordinary USB web camera offers enough video quality. While there is no agreement about the meaning of high-definition video, 720 scan lines (1280×720 pixels) is nowadays labelled as HD. The less quality Youtube offers in playing videos is 640×360, which is enough for many cases. A typical €20 USB web camera offers HD quality. The required electronic components and motors for making a tracking system supporting a camera are not expensive too. Tracking a person from RGB images is not trivial, though.

3D sensors are available for some time now. Only recently they have been available at affordable costs. 3D sensors are interesting for person tracking applications, because the third dimension facilitates the identification of the person. The Kinect ((4) and references therein) is a worthy representative of a non expensive, new class of 3D sensors suitable for person tracking applications. Additionally to easing the tracking task, it comes with an Application Programming Interface (API) that eases further such applications. Indeed, by making available the 3D coordinates of the body joints, depicted by the filled circles in Figure 1, one can add interesting features as are gesture recognition.

The project presented in this paper is about tracking the position of a person – the lecturer, and recording its movements in real-time (at no less than 25fps) on video (including sound). Explaining the system will take the rest of the paper. In section 2 we present the general panorama of lesson recording systems, and some attempts at tracking persons with 3D sensors, specifically with the Kinect sensor. We present our system in section 3 and the results of using it in section 4. Finally, in section 5, we present some conclusions, mentioning some possible developments.

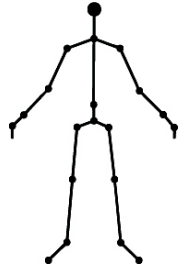

Figure 1- A skeleton as calculated by the Kinect's API.

## 2. CLASS RECORDING SYSTEMS – STATE OF THE ART

Given the interest in recording classes, as presented in the introduction, systems where the recording is made from a live classroom tend to impose a human involvement in the form of the logistics of pre and post recording, while the recording phase requires the presence of a technician (a camera operator). These tasks add up to the costs of universities, and interested institutions alike. (5) gives an idea of the costs of some systems. (6) and (7) give an account of the workload required by an institution to set up and maintain a class recording system. This by itself justifies the need for an automatic recording system.

Aware of this difficulty from universities, some companies developed full products they now sell. Systems as Tegrity (8), Echo36 0(9), and Panopto (10), to mention just a few, are not oriented to the recording phase, but downstream to that; they offer cloud services ranging from videos (prepared by the instructors) to their consumers (the students). (11) is presented as an equipment offering integration with room control systems and room scheduling platforms; regarding recording automation this is the farthest it goes. For key features of some products see (5).

Avoiding the good services of a camera operator typically means that the camera will be fixed. This setup in unacceptable given the current state of technology. Possibly one of the first, MScribe (12) is a robotic tracking system devoted to recording classes enforcing the use of a wearable device taking the form of an infrared-emitting necklace, "freeing the lecturer from being anchored to a podium and eliminating the operational costs of a camera operator" (from (12)). Such an active device facilitates a lot the task of tracking a person through an RGB camera, although it has notable disadvantages. The system may lose track of the speaker if he turns completely to the blackboard in order to write down some explanation. Also, for large lecture halls, the system may lose the lecturer's position while the speaker moves away from the camera.

The recent massification of 3D sensors allowed a new range of new applications. With respect to automating the recording of classes, as far as the authors knowledge goes, there are not many proposals. [5] uses a Kinect sensor to track the speaker's position. The Kinect sensor is fixed and tracks the position of the speaker, which it is then passed to a camera that follows the speaker, with horizontal movements (around the yaw axis). Additionally, it reacts to a single gesture that allows a snapshot to be taken, saving the time students need to take written notes.

The system is always focused on the lecturer, providing no detail of whatever subject is being projected on the white canvas. The horizontal movements of the camera are a bit restrictive. It would be better if the camera also moved vertically too (around the pitch axis). For example, if the camera is installed centrally near the front rows of students, when the lecturer approaches the students its head may be out of the camera's field of view, unless there is a tilt degree of freedom. Finally, having a fixed Kinect sensor limits the area of the lecture hall being covered. This is limiting, as the lecturer should not be constrained in its movements, on one hand, and also imposes limits on the size of the lecture hall in which the system can be applied, on the other hand.

## 3. PROJECT DESCRIPTION

In this section we describe the project; from analysis to construction.

### 3.1 Problem analysis and requirements

The system should be able to record on video a class automatically, i.e., without the presence of a cameraman. This requires the assignment of the cameraman role to the speaker itself. Avoiding the presence of a cameraman is a real (economical) gain for the institution, but also for the video watcher. In fact, in many recorded lessons available on the web, there are many examples of a cameraman that constantly follows the movements of the speaker by its camera, without paying attention to the specifics of the presentation, thus missing what is happening on the blackboard or the projection canvas, where at many times the most important part of the lesson is (e.g., a graphic, an equation, a table, etc.), from the point of view of the learner.

The role of video editor/director is then assigned to the speaker, meaning that he must decide what scene shall be recorded: him, the blackboard or the white canvas. The Kinect sensor (rich) features facilitated the decision about what type of Human-Machine interface to adopt. The use of the 3D coordinates of the body joints of the speaker serves two purposes in this application: first, it allows the speaker's position to be tracked (thus giving him more freedom to move around the classroom); second, makes obvious the choice of a gesture interface to command the system because it is practical.

Note that the video resolution of the Kinect sensor (640×480) is too low for the purpose of recording a class on video. Moreover, the Kinect sensor must always be focused on the speaker, in order not to miss any gesture he might perform. This creates the necessity of a total of two, independent, articulated mechanical systems, both having two degrees of freedom (azimuth and elevation): one supporting the Kinect sensor, which always tracks the speaker, and the other supporting the video camera, which records the scene as decided by the speaker.

Next we describe the two natural parts of the system: the hardware part, composed by the two mechanical systems and the respective electronic controllers; and the software part, running on a Personal Computer (PC), where the most interesting task is gesture recognition.

## 3.2 Gesture interface

Our first approach to the user interface, allowing the speaker to interact with the system, was to made available the images as captured by the Kinect sensor for the speaker to see during its presentation, and conceive its right hand as a mouse pointer. This requires the recognition of the speaker's right hand, which is not difficult, as the Kinect API provides the 3D coordinates of every body joint. As soon as the right hand moves away from the body, a symbol with a black hand is drawn over the hand, representing the fact that the hand can function as the mouse pointer. The speaker should then move in order for its right hand to reach the desired button in the image. When he does, the button background colour changes to green, as illustrated in Figure 2**Erro! A origem da referência não foi encontrada.**. Keeping its hand over the button for about two seconds means pressing the button, thus triggering a command.

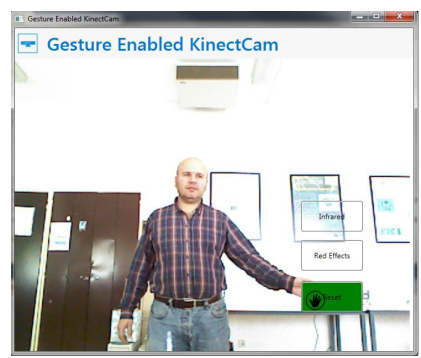

Figure 2- Illustration of the speaker's right hand seen as a mouse pointer. Note also three transparent buttons in the right side of the image. The buttons are painted with a solid color (in this case green) signaling that the button is about to be pressed.

This interface turned out not to be very practical. A few reasons attended to this. The speaker might be a little bit away from the desired button; thus having to move in order to approach the button, distracting himself from its line of thought. Also, when too close, the speaker might hit a button inadvertently and, making things worse, without even realising it. Finally, from the point of view of hardware, a system like this requires the presence of a screen somewhere between the audience and the speaker, which might occlude one from the other, besides the fact that rises the cost of the solution.

In reaction to this shortcoming, we decided to create a different gesture interface. We opted to recognize static gestures of the speaker's hands, i.e., gestures where the relative positions of both hands assume singular values that facilitate recognition. By static it is meant that both hands stay in the desired position for a fixed time interval.

Before defining each gesture, we present the commands each gesture represents. We listed four commands for the interface of the speaker with the system:

- **Start/stop recording**. The speaker uses this gesture to toggle the state of the recording of the presentation to file. This is important in order to create videos that have no uninteresting pieces of the class (e.g., between the start of the class and the beginning of the presentation). Also, it provides the means for the speaker to create video segments of the whole class. Remember that the speaker knows best when to segment the class video.

- **Record the canvas**. The speaker uses this command to have the video camera move to and focus the white canvas. He may choose this scene when it is best for the video watcher to focus on some technical aspect projected on the canvas (a formula, etc.).
- **Record the blackboard**. The same as the last point, but applied to the blackboard.
- **Record the speaker**. This command is triggered by the speaker when he decides that it is best for the video watcher to focus on the speaker, just because watchers like to focus on who is talking, or because he would like to complement its explanations with some form of hand animation, for instance.

The pictures of the postures the speaker must mimic in order to trigger the commands are presented in Figure 3, Figure 4, Figure 5 and Figure 6.

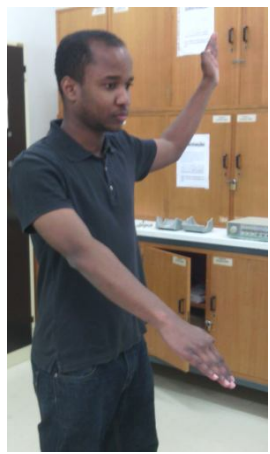

Figure 3 - Body posture for the "start/stop recording" command.

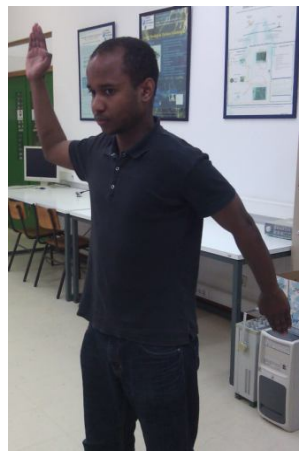

Figure 4- Body posture for the "record the white canvas" command.

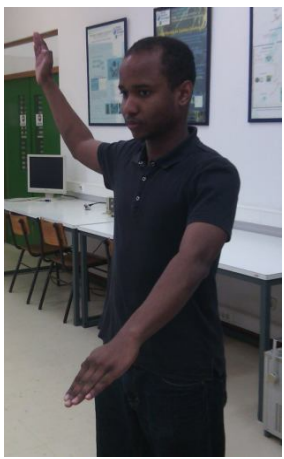

Figure 5- Body posture for the "record the blackboard" command.

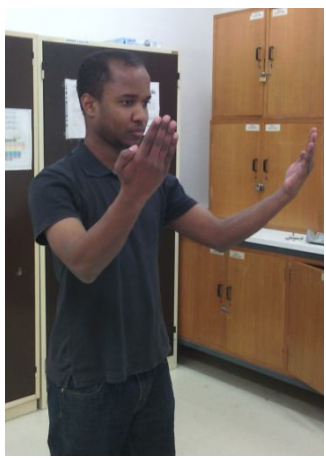

Figure 6- Body posture for the "record the speaker" command.

Gesture classification is based on the relative position of both wrists. In order to facilitate the evaluation of the wrist's location by the Kinect API, it is mandatory that, i) wrists don't suffer from occlusion, and ii) poses where two or more body articulations are collinear and aiming to the 3D sensor should be avoided.

As a general principle, gestures must be designed in order to maximize the distance between homologous writs, as a way to maximize gesture recognition and classification, while at the same time not requiring the speaker to perform gestures he might consider potentially embarrassing (specially in front of young students); as an example, we agreed on not having the speaker with arms stretched sideways. From the observation of the body poses in Figure 3 to Figure 6, the poses in Figure 4 and Figure 5 are the ones that are most similar, as the right wrist occupies the same position in space. In practice, this choice provoked no degradation in the recognition of the two corresponding gestures (more on this later).

Complementary to the relative positions of both wrists, a gesture is recognized if the corresponding spatial conditions are recognized for an uninterrupted period between two to four seconds. Because the speaker must maintain the gesture conditions during this time interval, the gestures are of the static kind. In relation to dynamic gestures, i.e., gestures that require movement of the wrist(s), static ones are easier to program and, in our opinion, more appropriate for this application.

## 3.2 System architecture

We now present the general system architecture in the form of a block diagram in Figure 7, and a picture of the whole system in Figure 8, except for the board containing the electronics.

From the analysis presented in sec. 3.1 (third paragraph), we need to construct two electromechanical systems both having pitch and yaw degrees of freedom. We call S1 the system to which the Kinect sensor is attached, and S2 the system carrying the video recorder. The kinect sensor is already composed by a motor drive allowing rotating movements allowing one to control its pitch. We then had to install a (stepper) motor to provide pan (or yaw) movements. The built prototype is presented in Figure 9.

The electromechanical system carrying the video camera, S2, was built from a off-the-shelf product - Merit LILIN PIH-302 outdoor pan/tilt - which we found in our stock of equipment. It is composed of two step motors which we decide to control directly.

In the diagram in Figure 7, an Arduino Uno is distinguished to be in between a personal computer and the motor drive electronics. Not present in the diagram is a set of five LEDs which provide visual feedback to the speaker about the state of the system (power is on, system is receiving a valid command during a two to four seconds window, S2 follows the speaker, S2 focus is on the blackboard, and S2 focus is on the white canvas).

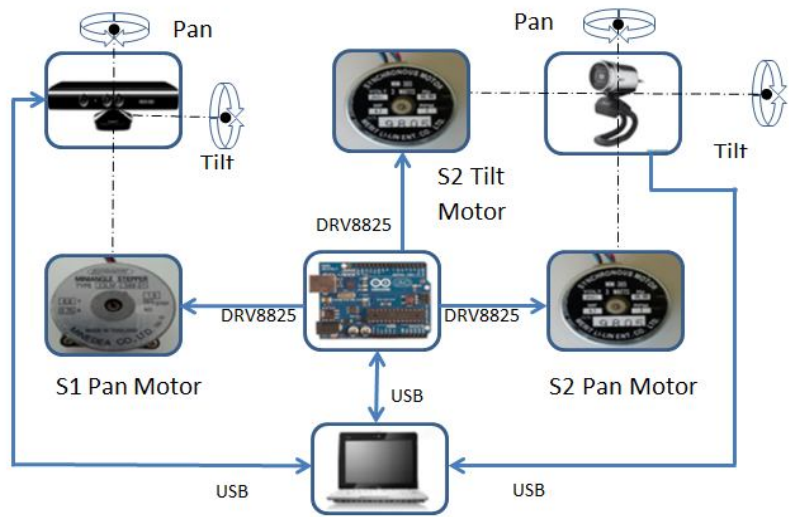

Figure 7 - Block diagram of the whole system. There are two electromechanical systems: S1, to which the Kinect sensor is fixed, and S2, to which the USB camera is fixed. S1 has two degrees of freedom: one is provided by the electrical motor inside the Kinect, supplying tilt control; the other is provided by a step motor providing pan control to the Kinect's field of view. The USB camera has also two degrees of freedom: pan and tilt.

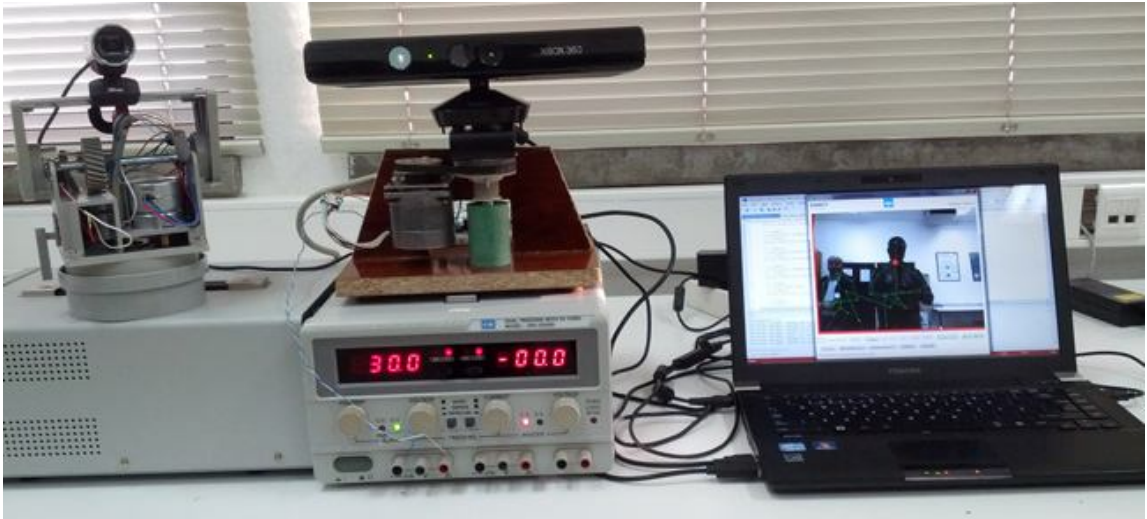

Figure 8 - Picture of the parts of the system, except for the electronics. From left to right: S2, S1 on top of a power supply, and a personal computer running the program where one can see the typical green segments the Kinect's API uses to draw the skeleton.

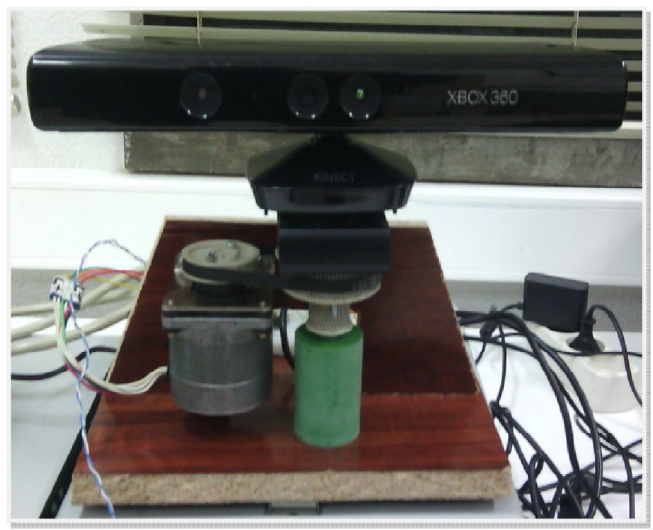

Figure 9 - View of the electromechanical system (denoted S1) to which the Kinect sensor is attached. The green cylinder acts as a support for the sensor. The step motor is the gray cylinder at the left of the support. The mechanical coupling between the motor and the sensor is made through a black belt linking two white pulleys.

A final topic regards the smooth control of acceleration of the movements of both S1 and S2. S1 (see Figure 9) carries the Kinect sensor; not providing smooth changes in velocity to S1 degrades the processing of the API functions (e.g., the 3D coordinates of the body joints are more prone to calculation errors). S2 (see Figure 10) carries the video recorder; not providing smooth movements to it implies recording sudden movements, degrading the class visualization experience.

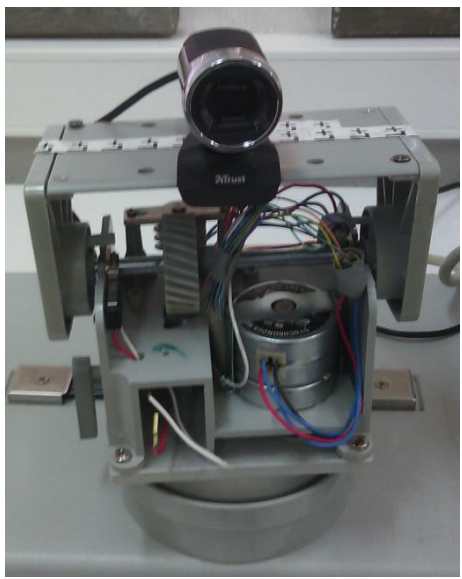

We implemented such a smooth behavior through a state machine (see Figure 11). We recurred to lookup tables and timer interrupts to implement acceleration and deceleration profiles that follow a sigmoid behavior. This approach proved to be correct; since there are three step motors to control, so efficiency is mandatory.

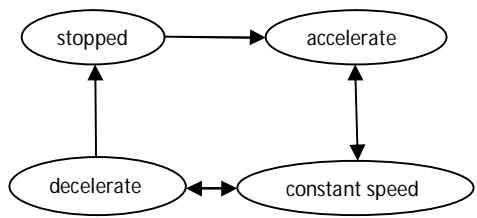

Figure 11 - state machine for changing speed smoothly.

### 3.3. Tests

In order to evaluate the performance of the system, we tested it under conditions close to a class in a real classroom. The gestures to test are: start the recorder, focus on the speaker, focus on the blackboard, focus on the white canvas, stop the recorder, following Figure 3 to Figure 6, and a delusive gesture (in fact, a small collection of gestures that are different from the defined ones, but share some similarities with them, in particular they were static and performed during about three seconds). The classroom has dimensions $6 \times 10$ meter. The system was placed centrally about 4 meters away from the front wall (i.e., the wall where the blackboard and white canvas are installed). Light was artificial. The test lasted for about 30 minutes, where the speaker moved normally as he does when giving a real class, sometimes gesticulating randomly, but normally (i.e., not too fast, not too slow). During the experiment, 20 instances of each gesture were performed, including the delusive one, intermingled with no special order, in a total of 120 instances (6 different gestures $\times$ 20 repetitions). The resulting confusion matrix is presented in Table 1.

Table 1: Confusion matrix for 20 instances of each of the five gestures, plus a delusive one, in a total of 120 performed gestures

| Gesture | Start | Stop | Blackboard | Canvas | Speaker | Undefined | Sum |
|---|---|---|---|---|---|---|---|
| Start | 18 | 0 | 0 | 0 | 0 | 2 | 20 |
| Stop | 1 | 18 | 0 | 0 | 0 | 1 | 20 |
| Blackboard | 0 | 0 | 20 | 0 | 0 | 0 | 20 |
| Canvas | 0 | 0 | 1 | 18 | 0 | 1 | 20 |
| Speaker | 0 | 0 | 0 | 0 | 20 | 0 | 20 |
| Undefined | 1 | 0 | 1 | 0 | 1 | 17 | 20 |
| Sum | 20 | 18 | 22 | 18 | 21 | 21 | 120 |

In Table 2 we present the success rate per gesture. For instance, the first line tells us that in 20 instances of Start, 90% were recognized as such. This rate is known as the true positive rate.

Table 2: Success rate per gesture

|  | Sucess Rate |
|---|---|
| Start | 90,0% |
| Stop | 90,0% |
| Blackboard | 100,0% |
| Canvas | 90,0% |
| Speaker | 100,0% |
| Undefined | 85,0% |

## 4. Concluding remarks

A double electromechanical system capable of recording classes on video was presented in this paper. The system was assembled from common parts: step motors, a 3D sensor (in this case a Kinect), a USB web camera, a laptop PC and a simple microcontroller board (in this case an Arduino).

The system is easily transported and installed in almost any classroom or showroom. It needs a simple installation procedure prior to its use; the locations of the blackboard and the white canvas are the only parameters needed.

What distinguishes the system from common approaches is that a cameraman is not needed. The role of video editor/director is assigned to the lecturer itself, which decides which scene to record through arm gestures. This is a very practical user interface, possibly superior to voice, because i) sometimes the classroom is a noisy environment, and ii) the normal speech can be understood as voice commands, increasing the false positive rate. The proposed system lowers the total cost of recording classes on video when in comparison to the traditional ones where a cameraman is needed. Because of this, the system can be considered state of the art.

# REFERENCES

1. **McClure, Ann.** Lecture Capture: A Fresh Look. *http://www.universitybusiness.com/.* [Online] 4 2008. http://www.universitybusiness.com/article/lecture-capture-fresh-look.

2. Massive open online course. *http://en.wikipedia.org/.* [Online] 19 2 2014. [Cited: 22 2 2014.] http://en.wikipedia.org/wiki/Massive_open_online_course.

3. An Early Report Card on Massive Open Online Courses. *http://online.wsj.com/.* [Online] 8 10 2013. [Cited: 22 2 2014.] http://online.wsj.com/news/interactive/MOOCchrtPRINT?ref=SB10001424052702303759604579093400834738972.

4. Kinect. *http://en.wikipedia.org/.* [Online] 28 1 2014. [Cited: 22 2 2014.] http://en.wikipedia.org/wiki/Kinect.

5. **Ramaswami, Rama.** Capturing the Market. *http://campustechnology.com/.* [Online] 6 1 2009. [Cited: 22 2 2014.] http://campustechnology.com/Articles/2009/06/01/Lecture-Capture.aspx.

6. *The Use and Benefit of a Xbox Kinect based Tracking System in a Lecture Recording System.* **Engelbert, B., Greweling, C., and Morisse, K.** [ed.] Jan Herrington et al. Chesapeake, VA, USA : s.n., 2013. Proceedings of World Conference on Educational Multimedia, Hypermedia and.

7. **Martyn, Margaret.** Engaging Lecture Capture: Lights, Camera. . . Interaction! *http://www.educause.edu/.* [Online] 22 12 2009. [Cited: 22 2 2014.] http://www.educause.edu/ero/article/engaging-lecture-capture-lights-camera-interaction.

8. Tegrity Campus. *http://www.tegrity.com.* [Online] [Cited: 22 2 2014.] http://www.tegrity.com/product.

9. Echo360. *http://echo360.com/.* [Online] 22 2 2014. http://echo360.com/.

10. Panopto. *http://panopto.com/.* [Online] [Cited: 22 2 2014.] http://panopto.com/.

11. ROOM CAPTURE. *http://www.sonicfoundry.com.* [Online] [Cited: 22 2 2014.] http://www.sonicfoundry.com/mediasite/room-capture-rl-recorders.

12. MScribe project. *http://www.umich.edu.* [Online] [Cited: 22 2 2014.] http://www.umich.edu/~mscribe/.